\documentclass[%
  twocolumn,
 showpacs,
 showkeys,
 preprintnumbers,
 amsmath,amssymb,
 aps,
  pra,
  longbibliography,
  floatfix,
 ]{revtex4-1}

\usepackage{tikz}
\usepackage[breaklinks=true,colorlinks=true,anchorcolor=blue,citecolor=blue,filecolor=blue,menucolor=blue,pagecolor=blue,urlcolor=blue,linkcolor=blue]{hyperref}
\usepackage{graphicx}
\usepackage{url}

\usepackage{xcolor}
\usepackage{colortbl}

\usepackage{eurosym}

\begin{document}

\title{A note on the statistical sampling aspect of delayed choice entanglement swapping}


\author{Karl Svozil}
\affiliation{Institute for Theoretical Physics, Vienna
    University of Technology, Wiedner Hauptstra\ss e 8-10/136, A-1040
    Vienna, Austria}
\affiliation{Department of Computer Science, University of Auckland, Private Bag 92019,  Auckland 1142, New Zealand}
\email{svozil@tuwien.ac.at} \homepage[]{http://tph.tuwien.ac.at/~svozil}


\begin{abstract}
Quantum and classical models for delayed choice entanglement swapping by postselection of measurements are discussed.
\end{abstract}

\maketitle


\begin{table*}
\begin{ruledtabular}
\begin{tabular}{c|cccccc|cccccc|cccccc}

\#  & A1       &       E2       &       E3       &       B4     & c/p & E   & A1       &       E2       &       E3       &       B4     & c/p & E     & A1       &       E2       &       E3       &       B4     & c/p & E      \\

\hline
\noalign{\vskip 0.3mm}
1       &        \cellcolor{yellow!20}  0       &        \cellcolor{white}      1       &        \cellcolor{white}      0       &        \cellcolor{blue!20}    1       &        \cellcolor{white}      p       &        \cellcolor{white}      p3      &        \cellcolor{yellow!20}  0       &        \cellcolor{white}      1       &        \cellcolor{white}      0       &        \cellcolor{blue!20}    1       &        \cellcolor{white}      p       &        \cellcolor{white}      p3      &        \cellcolor{green!20}   0       &        \cellcolor[gray]{0.8}  1       &        \cellcolor[gray]{0.8}  0       &        \cellcolor{green!20}   1       &        \cellcolor[gray]{0.8}  c       &        \cellcolor[gray]{0.8}  o2      \\
2       &        \cellcolor{red!20}     0       &        \cellcolor[gray]{0.8}  1       &        \cellcolor[gray]{0.8}  1       &        \cellcolor{red!20}     0       &        \cellcolor[gray]{0.8}  c       &        \cellcolor[gray]{0.8}  e2      &        \cellcolor{red!20}     0       &        \cellcolor[gray]{0.8}  1       &        \cellcolor[gray]{0.8}  1       &        \cellcolor{red!20}     0       &        \cellcolor[gray]{0.8}  c       &        \cellcolor[gray]{0.8}  e2      &        \cellcolor{red!20}     0       &        \cellcolor[gray]{0.8}  1       &        \cellcolor[gray]{0.8}  1       &        \cellcolor{red!20}     0       &        \cellcolor[gray]{0.8}  c       &        \cellcolor[gray]{0.8}  e2      \\
3       &        \cellcolor{green!20}   0       &        \cellcolor[gray]{0.8}  1       &        \cellcolor[gray]{0.8}  0       &        \cellcolor{green!20}   1       &        \cellcolor[gray]{0.8}  c       &        \cellcolor[gray]{0.8}  o2      &        \cellcolor{green!20}   0       &        \cellcolor[gray]{0.8}  1       &        \cellcolor[gray]{0.8}  0       &        \cellcolor{green!20}   1       &        \cellcolor[gray]{0.8}  c       &        \cellcolor[gray]{0.8}  o2      &        \cellcolor{green!20}   0       &        \cellcolor[gray]{0.8}  1       &        \cellcolor[gray]{0.8}  0       &        \cellcolor{green!20}   1       &        \cellcolor[gray]{0.8}  c       &        \cellcolor[gray]{0.8}  o2      \\
4       &        \cellcolor{red!20}     0       &        \cellcolor[gray]{0.8}  1       &        \cellcolor[gray]{0.8}  1       &        \cellcolor{red!20}     0       &        \cellcolor[gray]{0.8}  c       &        \cellcolor[gray]{0.8}  e2      &        \cellcolor{yellow!20}  0       &        \cellcolor{white}      1       &        \cellcolor{white}      1       &        \cellcolor{yellow!20}  0       &        \cellcolor{white}      p       &        \cellcolor{white}      p4      &        \cellcolor{red!20}     0       &        \cellcolor[gray]{0.8}  1       &        \cellcolor[gray]{0.8}  1       &        \cellcolor{red!20}     0       &        \cellcolor[gray]{0.8}  c       &        \cellcolor[gray]{0.8}  e2      \\
5       &        \cellcolor{yellow!20}  0       &        \cellcolor{white}      1       &        \cellcolor{white}      1       &        \cellcolor{yellow!20}  0       &        \cellcolor{white}      p       &        \cellcolor{white}      p4      &        \cellcolor{red!20}     0       &        \cellcolor[gray]{0.8}  1       &        \cellcolor[gray]{0.8}  1       &        \cellcolor{red!20}     0       &        \cellcolor[gray]{0.8}  c       &        \cellcolor[gray]{0.8}  e2      &        \cellcolor{red!20}     0       &        \cellcolor[gray]{0.8}  1       &        \cellcolor[gray]{0.8}  1       &        \cellcolor{red!20}     0       &        \cellcolor[gray]{0.8}  c       &        \cellcolor[gray]{0.8}  e2      \\
6       &        \cellcolor{green!20}   0       &        \cellcolor[gray]{0.8}  1       &        \cellcolor[gray]{0.8}  0       &        \cellcolor{green!20}   1       &        \cellcolor[gray]{0.8}  c       &        \cellcolor[gray]{0.8}  o2      &        \cellcolor{green!20}   0       &        \cellcolor[gray]{0.8}  1       &        \cellcolor[gray]{0.8}  0       &        \cellcolor{green!20}   1       &        \cellcolor[gray]{0.8}  c       &        \cellcolor[gray]{0.8}  o2      &        \cellcolor{yellow!20}  0       &        \cellcolor{white}      1       &        \cellcolor{white}      0       &        \cellcolor{blue!20}    1       &        \cellcolor{white}      p       &        \cellcolor{white}      p3      \\
7       &        \cellcolor{blue!20}    1       &        \cellcolor{white}      0       &        \cellcolor{white}      0       &        \cellcolor{blue!20}    1       &        \cellcolor{white}      p       &        \cellcolor{white}      p1      &        \cellcolor{red!20}     1       &        \cellcolor[gray]{0.8}  0       &        \cellcolor[gray]{0.8}  0       &        \cellcolor{red!20}     1       &        \cellcolor[gray]{0.8}  c       &        \cellcolor[gray]{0.8}  e1      &        \cellcolor{blue!20}    1       &        \cellcolor{white}      0       &        \cellcolor{white}      0       &        \cellcolor{blue!20}    1       &        \cellcolor{white}      p       &        \cellcolor{white}      p1      \\
8       &        \cellcolor{green!20}   1       &        \cellcolor[gray]{0.8}  0       &        \cellcolor[gray]{0.8}  1       &        \cellcolor{green!20}   0       &        \cellcolor[gray]{0.8}  c       &        \cellcolor[gray]{0.8}  o1      &        \cellcolor{green!20}   1       &        \cellcolor[gray]{0.8}  0       &        \cellcolor[gray]{0.8}  1       &        \cellcolor{green!20}   0       &        \cellcolor[gray]{0.8}  c       &        \cellcolor[gray]{0.8}  o1      &        \cellcolor{blue!20}    1       &        \cellcolor{white}      0       &        \cellcolor{white}      1       &        \cellcolor{yellow!20}  0       &        \cellcolor{white}      p       &        \cellcolor{white}      p2      \\
9       &        \cellcolor{blue!20}    1       &        \cellcolor{white}      0       &        \cellcolor{white}      1       &        \cellcolor{yellow!20}  0       &        \cellcolor{white}      p       &        \cellcolor{white}      p2      &        \cellcolor{blue!20}    1       &        \cellcolor{white}      0       &        \cellcolor{white}      1       &        \cellcolor{yellow!20}  0       &        \cellcolor{white}      p       &        \cellcolor{white}      p2      &        \cellcolor{blue!20}    1       &        \cellcolor{white}      0       &        \cellcolor{white}      1       &        \cellcolor{yellow!20}  0       &        \cellcolor{white}      p       &        \cellcolor{white}      p2      \\
10      &        \cellcolor{red!20}     0       &        \cellcolor[gray]{0.8}  1       &        \cellcolor[gray]{0.8}  1       &        \cellcolor{red!20}     0       &        \cellcolor[gray]{0.8}  c       &        \cellcolor[gray]{0.8}  e2      &        \cellcolor{red!20}     0       &        \cellcolor[gray]{0.8}  1       &        \cellcolor[gray]{0.8}  1       &        \cellcolor{red!20}     0       &        \cellcolor[gray]{0.8}  c       &        \cellcolor[gray]{0.8}  e2      &        \cellcolor{yellow!20}  0       &        \cellcolor{white}      1       &        \cellcolor{white}      1       &        \cellcolor{yellow!20}  0       &        \cellcolor{white}      p       &        \cellcolor{white}      p4      \\
11      &        \cellcolor{blue!20}    1       &        \cellcolor{white}      0       &        \cellcolor{white}      1       &        \cellcolor{yellow!20}  0       &        \cellcolor{white}      p       &        \cellcolor{white}      p2      &        \cellcolor{green!20}   1       &        \cellcolor[gray]{0.8}  0       &        \cellcolor[gray]{0.8}  1       &        \cellcolor{green!20}   0       &        \cellcolor[gray]{0.8}  c       &        \cellcolor[gray]{0.8}  o1      &        \cellcolor{green!20}   1       &        \cellcolor[gray]{0.8}  0       &        \cellcolor[gray]{0.8}  1       &        \cellcolor{green!20}   0       &        \cellcolor[gray]{0.8}  c       &        \cellcolor[gray]{0.8}  o1      \\
12      &        \cellcolor{green!20}   0       &        \cellcolor[gray]{0.8}  1       &        \cellcolor[gray]{0.8}  0       &        \cellcolor{green!20}   1       &        \cellcolor[gray]{0.8}  c       &        \cellcolor[gray]{0.8}  o2      &        \cellcolor{green!20}   0       &        \cellcolor[gray]{0.8}  1       &        \cellcolor[gray]{0.8}  0       &        \cellcolor{green!20}   1       &        \cellcolor[gray]{0.8}  c       &        \cellcolor[gray]{0.8}  o2      &        \cellcolor{green!20}   0       &        \cellcolor[gray]{0.8}  1       &        \cellcolor[gray]{0.8}  0       &        \cellcolor{green!20}   1       &        \cellcolor[gray]{0.8}  c       &        \cellcolor[gray]{0.8}  o2      \\
13      &        \cellcolor{green!20}   0       &        \cellcolor[gray]{0.8}  1       &        \cellcolor[gray]{0.8}  0       &        \cellcolor{green!20}   1       &        \cellcolor[gray]{0.8}  c       &        \cellcolor[gray]{0.8}  o2      &        \cellcolor{green!20}   0       &        \cellcolor[gray]{0.8}  1       &        \cellcolor[gray]{0.8}  0       &        \cellcolor{green!20}   1       &        \cellcolor[gray]{0.8}  c       &        \cellcolor[gray]{0.8}  o2      &        \cellcolor{yellow!20}  0       &        \cellcolor{white}      1       &        \cellcolor{white}      0       &        \cellcolor{blue!20}    1       &        \cellcolor{white}      p       &        \cellcolor{white}      p3      \\
14      &        \cellcolor{blue!20}    1       &        \cellcolor{white}      0       &        \cellcolor{white}      0       &        \cellcolor{blue!20}    1       &        \cellcolor{white}      p       &        \cellcolor{white}      p1      &        \cellcolor{blue!20}    1       &        \cellcolor{white}      0       &        \cellcolor{white}      0       &        \cellcolor{blue!20}    1       &        \cellcolor{white}      p       &        \cellcolor{white}      p1      &        \cellcolor{blue!20}    1       &        \cellcolor{white}      0       &        \cellcolor{white}      0       &        \cellcolor{blue!20}    1       &        \cellcolor{white}      p       &        \cellcolor{white}      p1      \\
15      &        \cellcolor{red!20}     1       &        \cellcolor[gray]{0.8}  0       &        \cellcolor[gray]{0.8}  0       &        \cellcolor{red!20}     1       &        \cellcolor[gray]{0.8}  c       &        \cellcolor[gray]{0.8}  e1      &        \cellcolor{blue!20}    1       &        \cellcolor{white}      0       &        \cellcolor{white}      0       &        \cellcolor{blue!20}    1       &        \cellcolor{white}      p       &        \cellcolor{white}      p1      &        \cellcolor{red!20}     1       &        \cellcolor[gray]{0.8}  0       &        \cellcolor[gray]{0.8}  0       &        \cellcolor{red!20}     1       &        \cellcolor[gray]{0.8}  c       &        \cellcolor[gray]{0.8}  e1      \\
16      &        \cellcolor{yellow!20}  0       &        \cellcolor{white}      1       &        \cellcolor{white}      0       &        \cellcolor{blue!20}    1       &        \cellcolor{white}      p       &        \cellcolor{white}      p3      &        \cellcolor{green!20}   0       &        \cellcolor[gray]{0.8}  1       &        \cellcolor[gray]{0.8}  0       &        \cellcolor{green!20}   1       &        \cellcolor[gray]{0.8}  c       &        \cellcolor[gray]{0.8}  o2      &        \cellcolor{green!20}   0       &        \cellcolor[gray]{0.8}  1       &        \cellcolor[gray]{0.8}  0       &        \cellcolor{green!20}   1       &        \cellcolor[gray]{0.8}  c       &        \cellcolor[gray]{0.8}  o2      \\
17      &        \cellcolor{blue!20}    1       &        \cellcolor{white}      0       &        \cellcolor{white}      0       &        \cellcolor{blue!20}    1       &        \cellcolor{white}      p       &        \cellcolor{white}      p1      &        \cellcolor{blue!20}    1       &        \cellcolor{white}      0       &        \cellcolor{white}      0       &        \cellcolor{blue!20}    1       &        \cellcolor{white}      p       &        \cellcolor{white}      p1      &        \cellcolor{red!20}     1       &        \cellcolor[gray]{0.8}  0       &        \cellcolor[gray]{0.8}  0       &        \cellcolor{red!20}     1       &        \cellcolor[gray]{0.8}  c       &        \cellcolor[gray]{0.8}  e1      \\
18      &        \cellcolor{yellow!20}  0       &        \cellcolor{white}      1       &        \cellcolor{white}      0       &        \cellcolor{blue!20}    1       &        \cellcolor{white}      p       &        \cellcolor{white}      p3      &        \cellcolor{yellow!20}  0       &        \cellcolor{white}      1       &        \cellcolor{white}      0       &        \cellcolor{blue!20}    1       &        \cellcolor{white}      p       &        \cellcolor{white}      p3      &        \cellcolor{green!20}   0       &        \cellcolor[gray]{0.8}  1       &        \cellcolor[gray]{0.8}  0       &        \cellcolor{green!20}   1       &        \cellcolor[gray]{0.8}  c       &        \cellcolor[gray]{0.8}  o2      \\
19      &        \cellcolor{blue!20}    1       &        \cellcolor{white}      0       &        \cellcolor{white}      1       &        \cellcolor{yellow!20}  0       &        \cellcolor{white}      p       &        \cellcolor{white}      p2      &        \cellcolor{green!20}   1       &        \cellcolor[gray]{0.8}  0       &        \cellcolor[gray]{0.8}  1       &        \cellcolor{green!20}   0       &        \cellcolor[gray]{0.8}  c       &        \cellcolor[gray]{0.8}  o1      &        \cellcolor{blue!20}    1       &        \cellcolor{white}      0       &        \cellcolor{white}      1       &        \cellcolor{yellow!20}  0       &        \cellcolor{white}      p       &        \cellcolor{white}      p2      \\
20      &        \cellcolor{green!20}   1       &        \cellcolor[gray]{0.8}  0       &        \cellcolor[gray]{0.8}  1       &        \cellcolor{green!20}   0       &        \cellcolor[gray]{0.8}  c       &        \cellcolor[gray]{0.8}  o1      &        \cellcolor{green!20}   1       &        \cellcolor[gray]{0.8}  0       &        \cellcolor[gray]{0.8}  1       &        \cellcolor{green!20}   0       &        \cellcolor[gray]{0.8}  c       &        \cellcolor[gray]{0.8}  o1      &        \cellcolor{blue!20}    1       &        \cellcolor{white}      0       &        \cellcolor{white}      1       &        \cellcolor{yellow!20}  0       &        \cellcolor{white}      p       &        \cellcolor{white}      p2      \\
21      &        \cellcolor{yellow!20}  0       &        \cellcolor{white}      1       &        \cellcolor{white}      1       &        \cellcolor{yellow!20}  0       &        \cellcolor{white}      p       &        \cellcolor{white}      p4      &        \cellcolor{yellow!20}  0       &        \cellcolor{white}      1       &        \cellcolor{white}      1       &        \cellcolor{yellow!20}  0       &        \cellcolor{white}      p       &        \cellcolor{white}      p4      &        \cellcolor{yellow!20}  0       &        \cellcolor{white}      1       &        \cellcolor{white}      1       &        \cellcolor{yellow!20}  0       &        \cellcolor{white}      p       &        \cellcolor{white}      p4      \\
22      &        \cellcolor{red!20}     0       &        \cellcolor[gray]{0.8}  1       &        \cellcolor[gray]{0.8}  1       &        \cellcolor{red!20}     0       &        \cellcolor[gray]{0.8}  c       &        \cellcolor[gray]{0.8}  e2      &        \cellcolor{red!20}     0       &        \cellcolor[gray]{0.8}  1       &        \cellcolor[gray]{0.8}  1       &        \cellcolor{red!20}     0       &        \cellcolor[gray]{0.8}  c       &        \cellcolor[gray]{0.8}  e2      &        \cellcolor{yellow!20}  0       &        \cellcolor{white}      1       &        \cellcolor{white}      1       &        \cellcolor{yellow!20}  0       &        \cellcolor{white}      p       &        \cellcolor{white}      p4      \\
23      &        \cellcolor{green!20}   0       &        \cellcolor[gray]{0.8}  1       &        \cellcolor[gray]{0.8}  0       &        \cellcolor{green!20}   1       &        \cellcolor[gray]{0.8}  c       &        \cellcolor[gray]{0.8}  o2      &        \cellcolor{yellow!20}  0       &        \cellcolor{white}      1       &        \cellcolor{white}      0       &        \cellcolor{blue!20}    1       &        \cellcolor{white}      p       &        \cellcolor{white}      p3      &        \cellcolor{yellow!20}  0       &        \cellcolor{white}      1       &        \cellcolor{white}      0       &        \cellcolor{blue!20}    1       &        \cellcolor{white}      p       &        \cellcolor{white}      p3      \\
24      &        \cellcolor{red!20}     1       &        \cellcolor[gray]{0.8}  0       &        \cellcolor[gray]{0.8}  0       &        \cellcolor{red!20}     1       &        \cellcolor[gray]{0.8}  c       &        \cellcolor[gray]{0.8}  e1      &        \cellcolor{red!20}     1       &        \cellcolor[gray]{0.8}  0       &        \cellcolor[gray]{0.8}  0       &        \cellcolor{red!20}     1       &        \cellcolor[gray]{0.8}  c       &        \cellcolor[gray]{0.8}  e1      &        \cellcolor{red!20}     1       &        \cellcolor[gray]{0.8}  0       &        \cellcolor[gray]{0.8}  0       &        \cellcolor{red!20}     1       &        \cellcolor[gray]{0.8}  c       &        \cellcolor[gray]{0.8}  e1      \\
25      &        \cellcolor{yellow!20}  0       &        \cellcolor{white}      1       &        \cellcolor{white}      0       &        \cellcolor{blue!20}    1       &        \cellcolor{white}      p       &        \cellcolor{white}      p3      &        \cellcolor{yellow!20}  0       &        \cellcolor{white}      1       &        \cellcolor{white}      0       &        \cellcolor{blue!20}    1       &        \cellcolor{white}      p       &        \cellcolor{white}      p3      &        \cellcolor{yellow!20}  0       &        \cellcolor{white}      1       &        \cellcolor{white}      0       &        \cellcolor{blue!20}    1       &        \cellcolor{white}      p       &        \cellcolor{white}      p3      \\
26      &        \cellcolor{red!20}     0       &        \cellcolor[gray]{0.8}  1       &        \cellcolor[gray]{0.8}  1       &        \cellcolor{red!20}     0       &        \cellcolor[gray]{0.8}  c       &        \cellcolor[gray]{0.8}  e2      &        \cellcolor{yellow!20}  0       &        \cellcolor{white}      1       &        \cellcolor{white}      1       &        \cellcolor{yellow!20}  0       &        \cellcolor{white}      p       &        \cellcolor{white}      p4      &        \cellcolor{yellow!20}  0       &        \cellcolor{white}      1       &        \cellcolor{white}      1       &        \cellcolor{yellow!20}  0       &        \cellcolor{white}      p       &        \cellcolor{white}      p4      \\
27      &        \cellcolor{blue!20}    1       &        \cellcolor{white}      0       &        \cellcolor{white}      1       &        \cellcolor{yellow!20}  0       &        \cellcolor{white}      p       &        \cellcolor{white}      p2      &        \cellcolor{blue!20}    1       &        \cellcolor{white}      0       &        \cellcolor{white}      1       &        \cellcolor{yellow!20}  0       &        \cellcolor{white}      p       &        \cellcolor{white}      p2      &        \cellcolor{green!20}   1       &        \cellcolor[gray]{0.8}  0       &        \cellcolor[gray]{0.8}  1       &        \cellcolor{green!20}   0       &        \cellcolor[gray]{0.8}  c       &        \cellcolor[gray]{0.8}  o1      \\
28      &        \cellcolor{red!20}     1       &        \cellcolor[gray]{0.8}  0       &        \cellcolor[gray]{0.8}  0       &        \cellcolor{red!20}     1       &        \cellcolor[gray]{0.8}  c       &        \cellcolor[gray]{0.8}  e1      &        \cellcolor{red!20}     1       &        \cellcolor[gray]{0.8}  0       &        \cellcolor[gray]{0.8}  0       &        \cellcolor{red!20}     1       &        \cellcolor[gray]{0.8}  c       &        \cellcolor[gray]{0.8}  e1      &        \cellcolor{red!20}     1       &        \cellcolor[gray]{0.8}  0       &        \cellcolor[gray]{0.8}  0       &        \cellcolor{red!20}     1       &        \cellcolor[gray]{0.8}  c       &        \cellcolor[gray]{0.8}  e1      \\
29      &        \cellcolor{red!20}     1       &        \cellcolor[gray]{0.8}  0       &        \cellcolor[gray]{0.8}  0       &        \cellcolor{red!20}     1       &        \cellcolor[gray]{0.8}  c       &        \cellcolor[gray]{0.8}  e1      &        \cellcolor{red!20}     1       &        \cellcolor[gray]{0.8}  0       &        \cellcolor[gray]{0.8}  0       &        \cellcolor{red!20}     1       &        \cellcolor[gray]{0.8}  c       &        \cellcolor[gray]{0.8}  e1      &        \cellcolor{blue!20}    1       &        \cellcolor{white}      0       &        \cellcolor{white}      0       &        \cellcolor{blue!20}    1       &        \cellcolor{white}      p       &        \cellcolor{white}      p1      \\
30      &        \cellcolor{red!20}     0       &        \cellcolor[gray]{0.8}  1       &        \cellcolor[gray]{0.8}  1       &        \cellcolor{red!20}     0       &        \cellcolor[gray]{0.8}  c       &        \cellcolor[gray]{0.8}  e2      &        \cellcolor{yellow!20}  0       &        \cellcolor{white}      1       &        \cellcolor{white}      1       &        \cellcolor{yellow!20}  0       &        \cellcolor{white}      p       &        \cellcolor{white}      p4      &        \cellcolor{red!20}     0       &        \cellcolor[gray]{0.8}  1       &        \cellcolor[gray]{0.8}  1       &        \cellcolor{red!20}     0       &        \cellcolor[gray]{0.8}  c       &        \cellcolor[gray]{0.8}  e2      \\

$\vdots$&
$\vdots$&
$\vdots$&
$\vdots$&
$\vdots$&
$\vdots$&
$\vdots$&
$\vdots$&
$\vdots$&
$\vdots$&
$\vdots$&
$\vdots$&
$\vdots$&
$\vdots$&
$\vdots$&
$\vdots$&
$\vdots$&
$\vdots$&
$\vdots$

\end{tabular}
\end{ruledtabular}

\caption{(Color online) Three partitions of, or views on, one and the same data set A1, E2, E3, B4 created through 30 simulated runs of an experiment.
There are the two uncorrelated singlet sources A1--E2 and E3--B4, producing random $0-1$- or $1-0$-pairs of data.
The difference between the three partitions lies in the choice of how the data E2 and E3 are interpreted:
If E2--E3 is interpreted as coincidence measurements ``revealing'' their relational properties, indicated by $c$ and a gray background,
then A1 and B4 are characterized by their relational properties; in particular, by the even and odd parity,
indicated by green and red backgrounds, respectively.
If, on the other hand, E2 and E3 are interpreted as measurements of single events, indicated by $p$ and a white background,
then A1 and B4 are characterized by their separate pairs of outcomes,
indicated by light yellow and blue backgrounds, respectively.
\label{2016-sampling-t1} }
\end{table*}

\section{Quantum case}

This is a very brief reflection on the sampling aspect of a paper~\cite{peres-DelayedChoiceEntanglementSwapping}
on delayed choice for entanglement swapping~\cite{Zuk-1993-entanglementswapping}.
The basic idea of entanglement swapping is as follows: at first
two uncorrelated pairs of entangled two-state particles in a singlet state are produced independently.
Then from each one of the two different pairs a single particle is taken.
These two particles are subsequently subjected to a measurement of their relational (joint) properties.
Depending on these properties the remaining two particles (of the two particle pairs)
can be sorted into four groups in a manner
which guarantees that within each group the pairs of remaining particles are entangled.
That is, effectively, (within each sort group) the remaining particles, although initially produced independently,
become entangled.

More explicitly, suppose the particles in the first pair are labelled by 1 and 2, and in the second pair by 3 and 4, respectively.
In the following only pure states will be considered.
The wave function is given by a product of two singlet state wave functions
\begin{equation}
| \Psi \rangle
=
| \Psi^-_{1,2} \rangle
| \Psi^-_{3,4} \rangle
,
\label{2016-sampling-e-state}
\end{equation}
where
$
| \Psi^\pm_{i,j} \rangle
=
\left({1}/{\sqrt{2}}\right)
\left(
| 0_i 1_j \rangle
\pm
| 1_i 0_j \rangle
\right)
$
and
$| \Phi^\pm_{i,j} \rangle
=
\left({1}/{\sqrt{2}}\right)
\left(
| 0_i0_j \rangle
\pm
| 1_i1_j \rangle
\right)
$
are the states associated with the Bell basis ${\frak B}_1 =\{| \Psi^- \rangle , | \Psi^+ \rangle , | \Phi^- \rangle , | \Phi^+ \rangle \}$
(or, equivalently, the associated context), ``$0$'' and ``$1$'' refers to the quantum numbers of the particles,
and the subscripts indicate the particle number.
In addition, consider the product states
$| 00 \rangle$,
$| 01 \rangle$,
$| 10 \rangle$, and
$| 11 \rangle$,
forming another possible basis (among a continuum of bases) ${\frak B}_2 =\{| -- \rangle , | -+ \rangle , | +- \rangle , | ++ \rangle \}$ of,
or context in, four dimensional Hilbert space.

Associated with these eight unit vectors in ${\frak B}_1$ and ${\frak B}_2$
are the eight projection operators from the dyadic products
$\textsf{\textbf{E}}^{\psi} = | \psi  \rangle \langle \psi |$,
with $\psi$ running over the entangled and product states, respectively.

Notice, for the sake of concreteness, that these states and projection operators can be represented by the vector components
$| 0  \rangle = \begin{pmatrix}1 , 0\end{pmatrix}^T$
and
$| 1  \rangle = \begin{pmatrix}0 , 1\end{pmatrix}^T$,
respectively; but these representations will not be explicitly used here.

The product (\ref{2016-sampling-e-state}) is a sum of products of
the states of
the two ``outer'' particles (particle $1$ from pair $1$ \& particle $4$ from pair $2$) and
the two ``inner'' particles (particle $2$ from pair $1$ \& particle $3$ from pair $2$);
it can be recasted in terms of the two bases in two ways:
\begin{equation}
\begin{split}
| \Psi \rangle
=
\frac{1}{2}\left(
| \Psi^+_{1,4} \rangle
| \Psi^+_{2,3} \rangle
-
| \Psi^-_{1,4} \rangle
| \Psi^-_{2,3} \rangle \right. \\ \left.
+
| \Phi^+_{1,4} \rangle
| \Phi^+_{2,3} \rangle
-
| \Phi^+_{1,4} \rangle
| \Phi^+_{2,3} \rangle
\right)
\end{split}
\label{2016-sampling-e-stater1}
\end{equation}
in terms of the bell basis; and, in terms of the product basis by
\begin{equation}
\begin{split}
| \Psi \rangle
=
| 0_1 1_4 \rangle
| 1_2 0_3 \rangle
-
| 0_1 0_4 \rangle
| 1_2 1_3 \rangle \\
-
| 1_1 1_4 \rangle
| 0_2 0_3 \rangle
+
| 1_1 0_4 \rangle
| 0_2 1_3 \rangle
.
\end{split}
\label{2016-sampling-e-stater2}
\end{equation}

Suppose an
agent Alice is recording the ``outer'' particle $1$,
agent Bob is recording the ``outer'' particle $4$,
and agent Eve is recording the ``inner'' particles $2$ and $3$, respectively.
Suppose further that Eve is free to choose her {\em type} of experiment
-- that is, either by observing the context
$\textsf{\textbf{E}}^{--}$,
$\textsf{\textbf{E}}^{-+}$,
$\textsf{\textbf{E}}^{+-}$, and
$\textsf{\textbf{E}}^{++}$
associated with the product basis, exclusive or observing the context
$\textsf{\textbf{E}}^{\Psi^-}$,
$\textsf{\textbf{E}}^{\Psi^+}$,
$\textsf{\textbf{E}}^{\Phi^-}$,  and
$\textsf{\textbf{E}}^{\Phi^+}$,
corresponding to the Bell basis states.
As a consequence of Eve's choice
the resulting state on Alice's and Bob's end is either a projection onto some (non-entangled) product state
$| ++ \rangle$,
$| +- \rangle$,
$| -+ \rangle$,  and
$| -- \rangle$,
exclusive or onto some entangled Bell basis state
$| \Psi^- \rangle$,
$| \Psi^+ \rangle$,
$| \Phi^- \rangle$,   and
$| \Phi^+ \rangle$,
respectively.

Peres' idea was to augment entanglement swapping with delayed choice; even to the point that Alice and Bob record their particles first;
and let Eve later, by a delayed choice~\cite{Ma-2016-RevModPhys.88.015005}, decide the type of measurement she chooses to perform:
Eve may measure propositions either corresponding to the elements of the Bell basis ${\frak B}_1$,
or of the product basis
${\frak B}_2$.
In the first case, in some quantum Hocus Pocus way, ``entanglement
is produced {\it a posteriori}, after the entangled particles
have been measured and may even no longer exist~\cite{peres-DelayedChoiceEntanglementSwapping}.''

In order to obtain a clearer picture, let us observe that, while Eve can choose between the two contexts (or measurement bases)
${\frak B}_1$
or
${\frak B}_2$,
she has no control of the particular outcome -- that is,
according to the axioms of quantum mechanics, the concrete state in which she finds the particles $2$ \& $3$ occurs irreducibly random,
with probability $1/4$ for each one of the terms in
(\ref{2016-sampling-e-stater1}) and
(\ref{2016-sampling-e-stater2}).

This can be interpreted as yet another instance of the peaceful coexistence~\cite{timpson-2002,Seevinck:2010eb}
between relativity and quantum mechanics, mediated by parameter independence but outcome dependence of such events:
Eve wilfully chooses the parameters -- in this case the Bell basis ${\frak B}_1$ {\it versus} the product basis ${\frak B}_2$ --
but quantum mechanics, and in particular, the recordings of Alice and Bob, are insensitive to that.
Yet, Eve cannot in any way choose or stimulate the outcomes at her side, which quantum mechanics is sensitive to.
(Actually, if Eve could somehow manipulate the outcome
-- maybe by stimulated emission~\cite{svozil-slash} -- this would be another instance of faster-than-light quantum communication,
and possibly also the end of peaceful coexistence.)

Eve's task is twofold: (i) in communicating the {\em type} of measurement performed
(Bell state {\it versus} product state observables),
Eve tells Alice and Bob whether she samples an entangled or a product state; and
(ii) in communicating her concrete measurement outcome Eve informs Alice and Bob about the concrete entangled state they are dealing with.
For the sake of an example of a protocol sentence of Eve, consider this one:
{\em ``I decided to measure my $i$th set of two particles $2$ \& $3$ in the Bell basis,
and found the particles to be in the singlet state $| \Psi^-_{2,3} \rangle$
(so your state should have also been a singlet one, namely $| \Psi^-_{1,4} \rangle$; and your outcomes are the inverse of mine;
that is, $i,j\rightarrow [(i+1) \text{ mod } 2], [(j+1) \text{ mod } 2]$).}

Thereby Eve is not merely {\em sampling}, but also {\em partitioning} the table of Alice's and Bob's recordings -- both according to her one choice of context,
as well as through her measurement outcomes.
Already Peres addressed this issue by stating
``the point is that it is meaningless to assert that two particles are entangled without
specifying in which state they are entangled, just as it is meaningless to assert that a
quantum system is in a pure state without specifying that state~\cite{peres-DelayedChoiceEntanglementSwapping}.''

\section{Classical analogue}

For the sake of making explicitly what this means, consider a classical analogue, and study binary observables in one measurement direction only.
Classical singlet states have been defined previously~\cite{peres222}, but as long as effectively one-dimensional
(with respect to the measurement direction) configurations are considered
it suffices to consider pairs of outcomes ``$0_i-1_j$'' or ``$1_i-0_j$,''
where the subscripts refer to the particle constituents.
These product states satisfy the property that the observables of the particles constituting that singlet are always different.
The associated observables are either joint observables, or separable ones.

Already at this point, it could quite justifiably be objected that this is an improper model for quantum singlets,
as it implies that the two particles constituting the singlet have definite individual observable values.
In contrast, a singlet quantum state is solely defined in terms
of the {\em correlations} (joint probability distributions)~\cite{CambridgeJournals:1737068,everett,mermin:753},
or, by another term, the {\em relational properties}~\cite{zeil-99,Timpson2003441} among the quanta;
whereby (with some reasonable side assumptions such as non-contextuality)
the supposition that the quanta carry additional information
about their definite individual states leads to a complete contradiction~\cite{pitowsky:218,2015-AnalyticKS}.

Nevertheless, if one accepts this classical model with the aforementioned provisions,
it is possible to explicitly study the partitioning of joint outcomes as follows.
Consider a concrete list of possible outcomes of two uncorrelated singlets
-- note that, as per definition, the constituents forming each singlet are
(intrinsically, that is within each singlet) correlated;
but the two singlets are externally uncorrelated --
as tabulated in Table~\ref{2016-sampling-t1}.
This is an enumeration of simulated empirical data -- essentially binary observables --
which are interpreted by {\em assigning} or {\em designating} some properties of a subensemble,
thereby effectively {\em inducing} or {\em rendering} some other properties or features on the remaining subensemble.

What is important here is to realize that the data allow many views or interpretations.
Consequently, what is a property of the data is purely conventionalized and means relative.
The only ontology relates to the pairs of statistically independent singlets; how their constituents relate to each other is entirely epistemic.
To emphasize this, Peres could be quoted a third time by repeating that ``it is obvious that from the raw data collected by Alice and Bob it is possible to
select in many different ways subsets that correspond to entangled pairs. The only
role that Eve has in this experiment is to tell Alice and Bob how to select such a
subset~\cite{peres-DelayedChoiceEntanglementSwapping}.''
It is amusing to notice that Peres' entire abstract applies to the analogue situation just discussed
(but we refrain from repeating it here because of fear of copyright infringement).

What are the differences between the classical analogue and the quantum original?
In answering this question one can consult another paper by Peres~\cite{peres222} on the hypothetical (non-)existence of
counterfactuals (or, in Specker's scholastic terminology~\cite{specker-60}, {\it Infuturabilien}).
One of the most striking differences is the fact that classical configurations
allow a truth table (that is, physical properties) of the constituents of the singlets,
whereas hypothetical (counterfactual) truth tables associated with entangled quantum states,
when viewed at different directions or contexts,
in general do not; at least not statistically~\cite{peres222},
but also not on a per particle pair basis~\cite{pitowsky:218,2015-AnalyticKS}.

That is, if we analyse the Bell states sampled according to Eve's directives by Alice and Bob,
they will be not only correlated but also entangled;
in particular, particles in a sampled singlet state will perform like a singlet state produced from a common source~\cite{Ma-2012-entanglementswapping}.
In particular, their correlations, involving more than one measurement directions, violate Bell-type inequalities.

\section{Type of randomness}

There is also another important difference in the perception of randomness involved.
The randomness in the classical analogue resides in the (pseudo-)random creation of the two singlet pairs.

In quantum mechanics certain entangled states, such as the states in the Bell basis ${\frak B}_1$,
exclude the separate existence of single-particle observables.
Formally this is easily seen, as tracing out one particle (i.e., taking the partial trace with respect to this particle)
yields the identity density matrix for the other particle:
for instance,
$
{\rm Tr}_i
\left(
\vert \Psi^\pm_{i,j} \rangle  \langle \Psi^\pm_{i,j}  \vert
\right)
=  {\rm Tr}_i
\left(
\vert \Phi^\pm_{i,j} \rangle  \langle \Phi^\pm_{i,j}  \vert
\right)
=
{\rm Tr}_j
\left(
\vert \Psi^\pm_{i,j} \rangle  \langle \Psi^\pm_{i,j}  \vert
\right)
=  {\rm Tr}_j
\left(
\vert \Phi^\pm_{i,j} \rangle  \langle \Phi^\pm_{i,j}  \vert
\right)
=
\left(1/2\right){\Bbb I}_2
$.

This is a consequence of the fact that (under certain mild side assumptions such as non-contextual value definiteness)
a quantum state can only be value definite with respect to a single one proposition~\cite{2015-AnalyticKS}--
that is, the proposition corresponding to the state preparation, which in turn corresponds to a single direction,
and a unit vector in the
$2^n$-dimensional Hilbert space of $n$ 2-state particles.
(A generalization to particles with $k$ states per particles is straightforward.)
Relative to this single value definite proposition,
all other propositions corresponding to non-orthogonal vectors are indeterminate.
Zeilinger's Foundational Principle~\cite{zeil-99,Timpson2003441}
is a corollary of this fact, once an orthonormal basis system including the
vector corresponding to this determinate property is fixed:
it is always possible to define filters corresponding to equipartitions of basis states which are co-measurable
and resolve states corresponding to single basis elements~\cite{DonSvo01,svozil-2002-statepart-prl}.

As has already been mentioned earlier,
Schr\"odinger~\cite{CambridgeJournals:1737068},
was the first to notice that, as expressed by Everett~\cite{everett}, in general
``a constituent subsystem cannot be said to be in
any single well-defined state, independently of the remainder
of the composite system.''
The entire state of multiple quanta
can be expressed completely in terms
of correlations (joint probability distributions)~\cite{wootters-1990-localaccOQStates,mermin:753},
or, by another term, relational properties~\cite{zeil-99},
among observables belonging to the subsystems.
There is ``a complete knowledge
of the whole without knowing the state of any one part. That a thing can be in a definite
state, even though its parts were not~\cite{Bennett-IBM-03.05.2016}.''

Some have thus suggested that, upon ``forcing'' the ``measurement'' of such indeterminate observables
the ``outcomes'' allow one to obtain ``irreducible randomness.''
In theological terms, this is a {\em creatio continua;} quasi {\it ex nihilo}.
Indeed, this appears to be the canonical position at present.

I have argued~\cite{svozil-2003-garda,svozil-2013-omelette}
that in such cases a context translation takes place that is effectively mediated by the measurement apparatus.
In many cases this apparatus may be considered quasi-classical; with many degrees of freedom which are, for
all practical purposes (but not in principle), impossible to resolve.
Therefore, the forced single outcome reflects both the microstate of the ``measurement device'' as well as the ``object,''
whereby the cut between those two is purely conventional~\cite{svozil-2001-convention} and, in close analogy to statistical mechanics~\cite{Myrvold2011237}
means relative.

\section{Concluding remarks}

Pointedly stated any set of raw data from correlated sources, quantum or otherwise,
can be combined and (re-)interpreted in many different ways.
Any such way presents a particular view on, or interpretation of, these data.
There is no unique way of representation; everything remains means relative and conventional.

Temporal considerations are not important here, because no causation,
just correlations are involved.
This is not entirely dissimilar to what has already been pointed out by Born, ``there are deterministic
relations which are not causal; for instance, any time table or
programmatic statement~\cite{born-metaph-1950}.''

The difference between the sampling of quantum and classical system is the scarcity of information
encoded in entangled quantum states,
which carry relational information about joint properties of the particles involved (a property they share with their classical counterparts)
but do not carry information about their single constituents, as classical states additionally do.

\acknowledgments{
This work was supported in part by the European Union, Research Executive Agency (REA),
Marie Curie FP7-PEOPLE-2010-IRSES-269151-RANPHYS grant.

Responsibility for the information and views expressed in this article lies entirely with the author.

The author declares no conflict of interest.

}


%

\end{document}